\documentclass[10pt]{article}

\usepackage[utf8]{inputenc}
\usepackage[a4paper,lmargin=1.5in, rmargin=1.5in, footnotesep=1cm plus 4pt minus 4pt]{geometry}
\usepackage{setspace}
\usepackage{amsmath,amsthm,amssymb,esint}
\usepackage{mathtools, nccmath}
\allowdisplaybreaks
\usepackage{multirow}
\usepackage{sansmath}
\usepackage{pgfplots}
\pgfplotsset{compat = newest}
\usepackage{listings}
\usepackage{color}
\usepackage{placeins}
\usepackage{multicol}
\usepackage{multirow}
\usepackage{enumitem}
\usepackage{lipsum}
\usepackage{cases}
\usepackage{epstopdf}
\usepackage{float}
\usepackage{booktabs}
\usepackage{dirtytalk}
\usepackage{subcaption}
\usepackage{graphicx}

\usepackage{comment}

\usepackage[T1]{fontenc}
\usepackage{lmodern}
\usepackage{times}

\usepackage{sectsty}
\allsectionsfont{\sffamily\bfseries}
\subsubsectionfont{\sffamily\bfseries}

\usepackage[font=sf,labelfont=bf]{caption}

\usepackage[font=sf]{caption}

\newtheoremstyle{mystyle1}
  {}
  {}
  {\itshape}
  {}
  {\sffamily\bfseries}
  {{\sffamily\mdseries .}}
  { }
  {}
\theoremstyle{mystyle1}

\newtheoremstyle{mystyle2}
  {}
  {}
  {}
  {}
  {\itshape\sffamily}
  {\textsf{.}}
  { }
  {}
\theoremstyle{mystyle2}

\usepackage{hyperref}
\usepackage{xcolor}
\definecolor{rsrs}{RGB}{19, 59, 123}
\definecolor{myred}{rgb}{0.7, 0.0, 0.0}
\definecolor{mygreen}{rgb}{0.0, 0.3, 0.0}
\hypersetup{
    colorlinks,
    linkcolor={rsrs},
    citecolor={rsrs},
    urlcolor={rsrs}
}

\usepackage{tocloft}

\usepackage[nottoc]{tocbibind}

\usepackage{kantlipsum}

\usepackage[
    backend=biber,
    citestyle=authoryear-icomp,
    bibstyle=authoryear,
    natbib=true,
    maxbibnames=99,
    maxcitenames=2,
    giveninits=true,
    terseinits=true,
    uniquelist=false
]{biblatex}
\DefineBibliographyStrings{english}{%
    andothers = {\em et\addabbrvspace al.}
}
\DeclareNameAlias{author}{last-first}
\DeclareFieldFormat*{title}{#1}
\renewbibmacro{in:}{}

\citetrackerfalse
\DefineBibliographyStrings{english}{%
  mathesis = {MSc thesis},
}

\usepackage[title]{appendix}

\usepackage{authblk}

\usepackage[hang]{footmisc}
\usepackage[ruled, algo2e]{algorithm2e} 
\usepackage{algpseudocode}
\SetKw{KwInitialization}{Initialize}
\SetKw{KwInput}{Input:}
\SetKw{KwOutput}{Output:}

\usepackage{changepage}

\usepackage{booktabs}
\title{Exjobb}
\begin{document}

\title{ {\sffamily Probabilistic Pareto plan generation for semiautomated multicriteria radiation therapy treatment planning} }
\author[$*$\textsf{,}$\dagger$]{Tianfang Zhang}
\author[$\dagger$]{Rasmus Bokrantz}
\author[$*$]{Jimmy Olsson}
{
\affil[$*$]{Department of Mathematics, KTH Royal Institute of Technology, Stockholm SE-100 44, Sweden}
\affil[$\dagger$]{RaySearch Laboratories, Eugeniavägen 18, Solna, Stockholm SE-103 65, Sweden}
}
\date{\textsf{January 18, 2022}}
\maketitle

\begin{quote}
{\centering
\section*{Abstract}
}
\textit{Objective:} We propose a semiautomatic pipeline for radiation therapy treatment planning, combining ideas from machine learning--automated planning and multicriteria optimization (MCO). 

\noindent \textit{Approach:} Using knowledge extracted from historically delivered plans, prediction models for spatial dose and dose statistics are trained and furthermore systematically modified to simulate changes in tradeoff priorities, creating a set of differently biased predictions. Based on the predictions, an MCO problem is subsequently constructed using previously developed dose mimicking functions, designed in such a way that its Pareto surface spans the range of clinically acceptable yet realistically achievable plans as exactly as possible. The result is an algorithm outputting a set of Pareto optimal plans, either fluence-based or machine parameter--based, which the user can navigate between in real time to make adjustments before a final deliverable plan is created. 

\noindent \textit{Main results:} Numerical experiments performed on a dataset of prostate cancer patients show that one may often navigate to a better plan than one produced by a single-plan-output algorithm. 

\noindent \textit{Significance:} We demonstrate the potential of merging MCO and a data-driven workflow to automate labor-intensive parts of the treatment planning process while maintaining a certain extent of manual control for the user.
\newline
\begin{spacing}{0.9}
{\sffamily\small \noindent \textbf{Keywords:} Knowledge-based planning, multicriteria optimization, dose prediction, dose--volume histogram prediction, uncertainty modeling, dose mimicking.}
\end{spacing}
\end{quote}

\tolerance=1000

\section{Introduction}

Having achieved a broad range of promising results in recent years, the application of machine learning methods to biomedical engineering is today established as a prosperous subject of research \citep{park, siddique}. Within automated treatment planning for radiation therapy, while the direct prediction of machine parameters of an optimal or desired plan remains an intractably high-dimensional and nonlinear problem, data-driven methods based on a prediction--mimicking pipeline has helped in homogenizing the labor-intensive process of creating clinically satisfactory plans \citep{berry}. However, such a methodology has several fundamental drawbacks---the produced deliverable plan is, for example, highly dependent on the quality of the prediction, and it is in practice often close to, yet not sufficiently in line with, the clinician's preferences, entailing the need for further post-processing using manual tools \citep{cagni}. Whereas some previous work has been devoted to address the former point \citep{babier_importance, nilsson, zhang_dvhpred}, the latter is especially concerning in that it implies not only strict requirements on the protocol used for creating the training data, but also significant time spent on model commissioning, which often involves setting up manual objectives supplementing the dose mimicking problem and, in turn, undermines the purpose of automated planning in the first place. Instead of continuing on the direction of developing a fully automated method, in this paper, we propose a new semiautomatic treatment planning workflow in which the a treatment planner or clinician may optionally articulate their own preferences by navigating in real time between Pareto optimal plans, combining ideas from machine learning and multicriteria optimization (MCO).

Automated treatment planning using machine learning, also known as knowledge-based planning, generally concerns the automatic plan generation using knowledge extracted from historically delivered clinical treatment plans \citep{ge, hussein, ng, wang}. Usually, one uses available data to train a machine learning model to predict the parameters of some parameterized optimization problem such that the solution will, to the largest extent possible, correspond to a clinically satisfactory plan. While some work has been focused on predicting weights in a weighted-sum objective function \citep{boutilier}, the majority of recent work has been aimed at predicting achievable dose-related quantities---for example, spatial dose distributions and dose--volume histograms (DVHs)---and setting up a dose mimicking optimization problem to minimize the deviation from the values evaluated on the actual dose to those predicted. 

Of such a prediction--mimicking pipeline, the prediction part has been extensively investigated in the literature. Regarding spatial dose prediction, current state-of-the-art methods are mostly based on convolutional neural networks with a U-net--like architecture \citep{campbell, kearney, nguyen_unet, shiraishi, babier_gan, nilsson}. For DVH or dose statistic prediction, methods based on overlap volume histograms as inputs \citep{appenzoller, jiao, ma_features, wall}, some involving predicting principal component coefficients of DVHs \citep{yuan, zhu}, have been more recently joined by those using input images directly as input \citep{liu, nguyen_pareto, zhang_dvhpred}. As discussed by \citet{babier_importance} and furthermore demonstrated in a previous paper by us \citep{zhang_dvhpred}, a substantial disconnect between the halves of a prediction--mimicking division may emerge from making deterministic inferences due to typical dose mimicking formulations being relatively non-robust to prediction errors. In this context, we showed \citep{zhang_dvhpred} that probabilistic methods, which output predictive probability distributions expressing estimation uncertainties, may reduce the information loss between the prediction and mimicking parts. Indeed, much of other previous work \citep{nilsson, nguyen_bagging, fogliata, covele} have already been directed toward precise quantification of predictive uncertainties for spatial dose or DVH statistics.

Despite this, the sensitivity of the final result to prediction errors brings to light more fundamental weaknesses of the current automated treatment planning paradigm. To approach clinical quality of automatically produced plans, the dataset used for training the prediction model must be sufficiently large as well as highly adherent to the protocol and planning standards set for creating the plans \citep{cagni, vanderbijl}---however, even then, certain manual intervention in terms of additional objectives or post-processing is sometimes needed to approach satisfactory quality \citep{zhang_dvhpred, mcintosh_nature}. In fact, as noted by \citet{zhangge} and \citet{vanderbijl}, automatically produced plans are the product of a population-based decision made upon the planning tradeoffs without first estimating or exploring what they are. Considering the large prevalence of inter-planner variations even among experienced planners \citep{nelms}, it becomes questionable whether the substantial efforts required to collect large and high-quality datasets is motivated by the end result, particularly if the dose mimicking formulation is heavily controlled by domain-knowledge objectives whose parameters need to be manually tuned and commissioned. Thus, instead of trying to further improve parts of a single-plan-output pipeline, we will direct attention to an MCO-based alternative semiautomatic workflow in which the output is rather a range of plans around what would have been the single-plan output, optionally letting the user make final adjustments through a real-time navigation interface. In contrast to conventional MCO, such a methodology enjoys the benefits of not needing to manually specify appropriate tradeoffs and constraints and of the Pareto surface covered being significantly smaller, thus requiring fewer Pareto plans to obtain comparable discretization accuracy. While the idea of restricting a standard Pareto surface to clinically acceptable regions has been found to be successful \citep{serna, goli, huang}, in a machine learning context, recent efforts have been focused on exploring tradeoff directions in DVHs by residual modeling \citep{zhangge} or predicting the spatial dose distributions of plans Pareto optimal with respect to a pre-specified MCO problem \citep{vanderbijl, nguyen_pareto, jensen, bohara}. Such methods have the advantage of speed but the disadvantage of, again, being vulnerable to prediction errors---for instance, if one were to navigate between a set of predicted Pareto optimal doses, it is not clear why the interpolated dose should correspond to or even be close to a physically realizable dose, especially in the presence of prediction errors. 

In this paper, we propose a semiautomatic treatment planning pipeline in which knowledge extracted from historically delivered clinical plans is leveraged to produce for each new patient a set of Pareto optimal plans, enabling the possibility for the user to articulate their preferences before a final plan is decided upon. To achieve this, we combine ideas from previous work on data-driven methods in automated planning with the flexibility of MCO framework, estimating predictive probability distributions to be translated into tradeoff objective functions. In particular, apart from models predicting spatial dose and dose statistics neutrally, as a basis for the resulting MCO problem, we will also create biased---or, as we will call it, \emph{tilted}---versions of the models simulating the cases of having optimized more aggressively toward certain groups of planning goals. In particular, a standard three-dimensional convolutional U-net is used to predict spatial doses, trained using a combination of a Sobolev space--inspired spatial loss and a DVH loss. The dose statistic prediction, on the other hand, is performed probabilistically by the similarity-based Bayesian mixture-of-experts model proposed in \citet{zhang_sbmoe} and used in \citet{zhang_dvhpred}. Tilted predictions are subsequently constructed using a change of probability measure in the predictive distribution of dose statistics and retraining of the neutral U-net model using an accordingly modified DVH loss. Translating the neutral and tilted predictions into tradeoff objectives, the resulting MCO problem is solved numerically to produce a set of Pareto optimal plans to be presented to the user for navigation. 

Specifically, the proposed means of setting up a well-defined MCO problem enables the generation of, for example, fluence-based Pareto plans rather than merely predictions thereof, an idea which has not been previously explored in the literature. The result is a novel semiautomatic workflow in which the Pareto plans are automatically generated relatively quickly and, compared to conventional MCO, span more exactly the range of achievable as well as clinically acceptable plans, all while having relatively limited requirements on training data size and quality. Apart from the proposed pipeline itself, conceptual contributions of this paper include the loss functions used for training the spatial dose prediction model and the notion of tilting predictions to simulate a change of tradeoff preferences. Numerical experiments demonstrate that one may often navigate to a better plan than the single-plan output associated with the neutral predictions and that the corresponding deliverable plan is often better than the clinical ground truth, showcasing the potential of uniting classical MCO with the current data-driven treatment planning paradigm.

\section{Method}

Let $\mathcal{X}$ and $\mathcal{D}$ denote vector spaces of contoured CT images and spatial dose distributions, respectively, and let $\{(x^n, d^n)\}_{n=1}^N \subset \mathcal{X} \times \mathcal{D}$ be a training dataset of clinical, historically delivered treatment plans. For sets such as $\{s_i\}_{i=1}^I$ or $\{s_i\}_{i \in R}$, we will use the shorthand notation $\{s_i\}_i$ when the range of the index $i$ is known from context, and analogously for vectors with parentheses instead of brackets. Using a linear voxel indexing, we can represent each $x \in \mathcal{X}$ and $d \in \mathcal{D}$ as vectors $x = (x_i)_i$ and $d = (d_i)_i$ of equal length. Let $\mathcal{R} = \mathcal{R}_{\operatorname{target}} \cup \mathcal{R}_{\operatorname{OAR}}$ be the set of regions of interest (ROIs) defined for the treatment site, partitioned into targets $\mathcal{R}_{\operatorname{target}}$ and organs at risk (OARs) $\mathcal{R}_{\operatorname{OAR}}$. For our purposes, we will only use the binary encodings of the ROIs and not the radiodensities in the CT images. Also, let $\{\psi_j\}_j$ be a generic collection of dose statistics $\psi_j : \mathcal{D} \to \mathbb{R}$ containing all dose-related quantities of interest---for example, dose-at-volume $\operatorname{D}_v$ or lower/upper mean-tail-dose $\operatorname{MTD}^{\pm}_v$ \citep{romeijn} levels at different volumes across different ROIs. We will often use $y \in \mathcal{Y}$ for the vector $(\psi_j(d))_j$ of dose statistic values evaluated on a dose $d$, with $\mathcal{Y}$ being the corresponding vector space.

\begin{figure}[H]
\centering
\includegraphics[width=\textwidth]{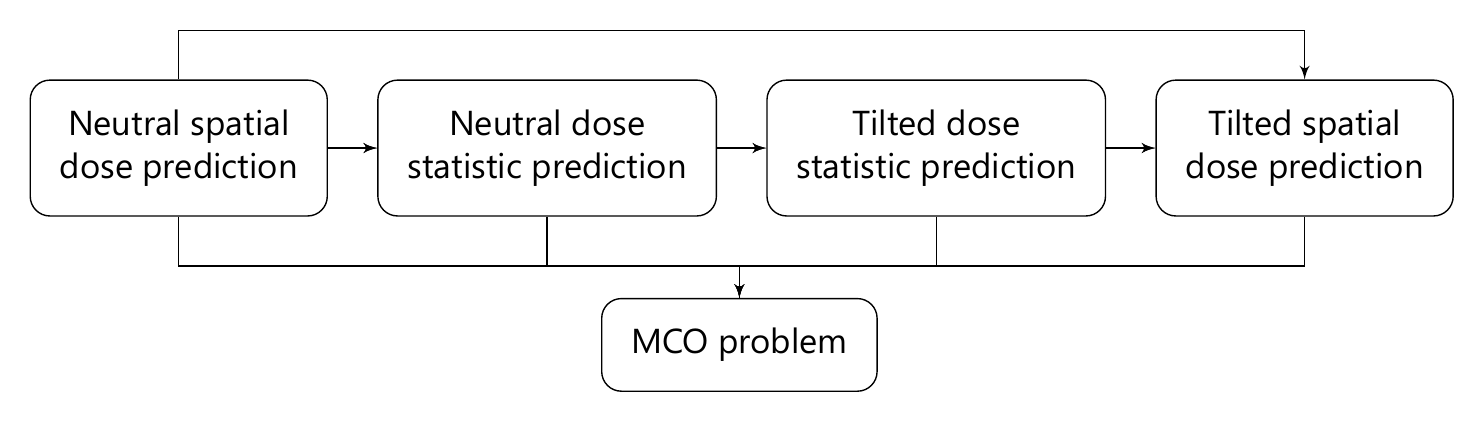}
\caption{Flowchart illustrating the different parts of the proposed pipeline.}
\label{pipeline}
\end{figure}

For an out-of-sample patient $x^* \in \mathcal{X}$, our task is to predict the corresponding spatial dose $d^* \in \mathcal{D}$ and its dose statistic values $y^* = (\psi_j(d^*))_j \in \mathcal{Y}$ and solve an appropriately constructed MCO problem to obtain a set of Pareto optimal plans. While the dose statistics are directly evaluable using the predicted spatial dose, it will be crucial to estimate the multivariate predictive distribution over groups of dose statistics rather than merely a single, deterministic prediction, which motivates our employing a separate dose statistic prediction model---the inter-statistic dependencies will help in creating tilted predictions, on which the MCO tradeoffs will be based. In particular, the proposed pipeline, shown in Figure \ref{pipeline}, will comprise the following parts:

\begin{enumerate}
\item a neutral spatial dose prediction model, estimating for each new patient $x^*$ the most likely dose distribution $d^*$;
\item a neutral dose statistic prediction model, estimating the multivariate predictive distribution $p(y^* \mid x^*, \{(x^n, y^n)\}_n)$, which includes feature extraction in both input and output spaces $\mathcal{X}$ and $\mathcal{Y}$;
\item a tilting method for the neutral dose statistic prediction model, which modifies the neutral prediction into ones that are \say{overly optimistic} in different groups of dose statistics;
\item a tilting method for the neutral spatial dose prediction model, which modifies the spatial dose prediction into ones more compatible with the tilted dose statistic predictions; and
\item the setup, numerical solution and navigation of an MCO problem based on the neutral and tilted predictions, where one tradeoff objective is constructed from each pair of spatial dose and dose statistic predictions.
\end{enumerate}

\subsection{Spatial dose prediction}
\label{doseprediction}

As a start, we need a spatial dose prediction model which can predict the dose distribution $d^*$ for each new patient $x^*$. Ideally, one would prefer to obtain the full predictive distribution $p(d^* \mid x^*, \{(x^n, d^n)\}_n)$, but although some previous work \citep{nilsson, nguyen_bagging} has been devoted to uncertainty estimation for spatial dose prediction, no existing method is able to output the complete multivariate predictive distribution as a closed-form probability density. Instead, we will proceed with a deterministic U-net dose prediction model $f : \mathcal{X} \to \mathcal{D}$ and an associated regression problem on the form
\[
d = f(x) + \epsilon
,\]
where $\epsilon = (\epsilon_i)_i$ is a non-isotropic regression error for which we assume the $\epsilon_i$ to be independent and normally distributed as $\epsilon_i \sim \operatorname{N}(0, \sigma_i^2)$---note, however, that the particular architecture used is not conceptually important. For each input $x = (x_i)_i \in \mathcal{X}$, we define the $i$th-position information $x_i$ as the $(|\mathcal{R}| + |\mathcal{R}_{\operatorname{target}}|)$-dimensional vector such that the first $|\mathcal{R}|$ components are the binary ROI encoding of voxel $i$ and the last $|\mathcal{R}_{\operatorname{target}}|$ components are distances to the targets. Explicitly, this is written as
\begin{equation}
\label{voxelinfo}
x_i = (1_{i \in R})_{R \in \mathcal{R}} \oplus (\operatorname{DT}_{R}(i))_{R \in \mathcal{R}_{\operatorname{target}}},
\end{equation}
where $\oplus$ denotes vector concatenation and $\operatorname{DT}_R(i)$, shorthand for distance transform, is the Euclidean distance from voxel $i$ to the voxel in $R$ nearest to $i$.  

For training, it is straightforward to see that a standard maximum-likelihood approach to fit the network weights in $f$ is equivalent to a weighted mean squared error minimization under the current assumptions on $\epsilon$, the weight of each voxel $i$ proportional to $\sigma_i^{-2}$. In addition to a mean squared error, we will introduce two additional loss contributions: one to account for the fact that a physical dose must vary smoothly in space, and one to enforce that the predicted DVHs are reasonable. For the first contribution, let $u : \mathbb{R}^3 \to \mathbb{R}$ be the three-dimensional scalar field representing a dose distribution $d$ in continuous space, and note that the standard mean squared error $|d - \hat{d}|^2$ in discrete space corresponds to the squared $L^2$-norm error $\Vert u - \hat{u} \Vert_{L^2}^2$ with $\Vert u \Vert_{L^2}^2 = \int u^2 \, dx$. Inspired by theory on partial differential equations, where one often works in Sobolev spaces \citep{evans}, we will instead use the Sobolev $H^1$-norm $\Vert u \Vert_{H^1}^2 = \int (u^2 + |\nabla u|^2) \, dx$. Translated into discrete space and adapting to the voxel weights arising from the non-isotropic likelihood, we write the spatial loss contribution $L_{\operatorname{spat}} : \mathcal{D}^2 \to \mathbb{R}$ as 
\begin{equation*}
L_{\operatorname{spat}}(d, \hat{d}) = \sum_i \frac{1}{\sigma_i^2} \left( (d_i - \hat{d}_i)^2 + \alpha |\nabla d_i - \nabla \hat{d}_i|^2 \right),
\end{equation*}
where $\alpha > 0$ controls the contribution of the gradient term and $\nabla d_i$ is used to denote the central difference approximation to the spatial gradient in the voxel grid at index $i$. While similar loss functions have been used in image reconstruction \citep{lu, vanderjeught} and meteorology \citep{hohlein} contexts, the idea is new to the dose prediction literature. In addition to the spatial loss, we will use a DVH-based loss $L_{\operatorname{DVH}} : \mathcal{Y}^2 \to \mathbb{R}$ similar to that proposed by \citet{nguyen_pareto}, but with squared differences taken in the dose direction rather in the volume direction. Specifically, if $\{\psi_j\}_{j \in S_{\operatorname{D}}}$ is a subset of all dose statistics restricted to only those of dose-at-volume type, we have 
\[
L_{\operatorname{DVH}}(y, \hat{y}) = \sum_{j \in S_{\operatorname{D}}} \frac{1}{\sigma_j^2} (y_j - \hat{y}_j)^2
,\]
where the $\sigma_j$ are introduced to entail a weighting analogously as the $\sigma_i$. Combining the spatial and DVH-based losses, the loss function $L : \mathcal{D}^2 \to \mathbb{R}$ used for training is written as
\[
L(d, \hat{d}) = L_{\operatorname{spat}}(d, \hat{d}) + L_{\operatorname{DVH}}\!\left( (\psi_j(d))_j, (\psi_j(\hat{d}))_j \right)
,\]
where the total loss to be minimized is given by the mean $N^{-1} \sum_{n=1}^N L(y^n, f(x^n))$ over the training examples. To ensure differentiability of each $\psi_j(d)$ with respect to $d$, for each forward pass in the neural network, the local dose $(d_i)_{i \in R}$ in each ROI $R$ is evaluated, sorted and linearly interpolated to be able to approximate the dose-at-volume value $\operatorname{D}_v(d)$ in $R$ for each $0 \leq v \leq 1$.

\subsection{Dose statistic prediction}

When predicting the dose statistic values $y^* = (\psi_j(d^*))_j$ given $x^*$, we now seek a probabilistic method to estimate the multivariate predictive density $p(y^* \mid x^*, \{(x^n, y^n)\}_n)$. These are essentially the same preconditions---high-dimensional inputs and outputs compared to data size, complex input--output relations and possibly skewed distributions---that motivated the development of the dose statistic prediction method using a mixture-of-experts model in \citet{zhang_dvhpred}, with the exception that a spatial dose prediction method has already been trained. Thus, instead of constructing and training a variational autoencoder for feature extraction as in \citet{zhang_dvhpred}, we will use the vector $(\psi_j(f(x)))_j$ of dose statistics evaluated on the predicted dose $f(x)$ as inputs to the mixture-of-experts model. This will be combined with purely geometric features $\phi_{\operatorname{geom}}(x)$, such as the discretized target distance transforms used for comparison in \citet{zhang_dvhpred}, to produce for each input $x$ a total feature vector
\begin{equation}
\label{dspredinput}
\phi_{\operatorname{tot}}(x) = \phi_{\operatorname{geom}}(x) \oplus (\psi_j(f(x)))_j.
\end{equation}
To further regularize the input space by reducing its dimension, which will be beneficial for the subsequent mixture-of-experts model, we will fit an isomap transformation \citep{tenenbaum} to the total feature vectors $\{\phi_{\operatorname{tot}}(x^n)\}_n$, creating corresponding embeddings $\{\phi_{\operatorname{iso}}(x^n)\}_n$. The isomap is a generic dimensionality reduction algorithm producing embedding vectors for which similarities are based on geodesic rather than Euclidean distances, which suits our purposes well. Similarly, for the output space $\mathcal{Y}$, we will extract the main principal components of the centered data $\{y^n - \overline{y}\}_n$, where $\overline{y}$ is the sample mean, writing $y^n = \overline{y} + P y_{\operatorname{pc}}^n$ for each $n$ with the principal components as columns in $P$ and the $y_{\operatorname{pc}}^n$ being coefficient vectors. Here, as we shall see below, it is important that each $y$ may be reconstructed as an affine transformation of its corresponding coefficient vector $y_{\operatorname{pc}}$. 

Having obtained a preprocessed dataset $\{(\phi_{\operatorname{iso}}(x^n), y_{\operatorname{pc}}^n)\}_n$, we can proceed by applying the same similarity-based mixture-of-experts model \citep{zhang_sbmoe} as used in \citet{zhang_dvhpred}, which outputs predictive distributions as multivariate Gaussian mixtures. This is a recently developed nonparametric Bayesian regression method in which predictions are based on input similarities rather than explicitly modeling input--output relations. With experts as multivariate normal distributions $\{\operatorname{N}(\mu_c, \Sigma_c)\}_{c=1}^C$, for each new input $\phi_{\operatorname{iso}}(x^*)$, mixture weights are calculated by first evaluating the similarity $\tau_n$ to each training input $\phi_{\operatorname{iso}}(x^n)$, and then the probability $\sigma_{nc}$ of each training output $y_{\operatorname{pc}}^n$ belonging to each expert class $c$---the total mixture weight $\pi_c$ for each expert class $c$ is then given by $\pi_c = \sum_n \tau_n \sigma_{nc}$. In particular, input similarities are based on the Mahalanobis distances $(\phi_{\operatorname{iso}}(x) - \phi_{\operatorname{iso}}(x'))^{\operatorname{T}} \Lambda (\phi_{\operatorname{iso}}(x) - \phi_{\operatorname{iso}}(x'))$, where $\Lambda$ is a precision matrix. The model has a predictive likelihood on the form
\[
p(y_{\operatorname{pc}}^* \mid \phi_{\operatorname{iso}}(x^*), \{(\phi_{\operatorname{iso}}(x^n), y_{\operatorname{pc}}^n)\}_n, \theta) = \sum_{c} \pi_c \operatorname{N}(y_{\operatorname{pc}}^* \mid \mu_c, \Sigma_c)
,\]
where $\theta = (\Lambda, \{(\mu_c, \Sigma_c)\}_c)$ are the model parameters and where
\[
\tau_n \propto \operatorname{N}(\phi_{\operatorname{iso}}(x^*) \mid \phi_{\operatorname{iso}}(x^n), \Lambda^{-1}), \quad \sigma_{nc} \propto \operatorname{N}(y_{\operatorname{pc}}^n \mid \mu_c, \Sigma_c)
\]
---for further details, see \citet{zhang_sbmoe} and \citet{zhang_direct}. Ultimately, the original predictive likelihood (and thus also the corresponding Monte Carlo--sampled predictive distribution) is obtained from the relation $y^* = \overline{y} + P y_{\operatorname{pc}}^*$ again as a Gaussian mixture
\[
p(y^* \mid x^*, \{(x^n, y^n)\}_n, \theta) = \sum_c \pi_c \operatorname{N}(y^* \mid \overline{y} + P\mu_c, P \Sigma_c P^{\operatorname{T}})
.\]

\subsection{Tilting}

With the neutral spatial dose and dose statistic prediction models in place, in order to be able to construct an MCO problem rather than a single-plan dose mimicking problem, our next step is to tilt the predictions to be more aggressive toward one or several planning goals. More precisely, whereas the neutral prediction may be associated with a balanced prioritization of the planning goals used in the historical plans, we wish to predict the outcome had the prioritization been heavily biased toward certain goals---for example, an extra low rectum dosage for a prostate plan, possibly at the cost of sacrificing target coverage---without actually needing to include such plans in the training data. On the other hand, in contrast to conventional MCO, we also wish to exclude completely unrealistic plans---for example, one with unacceptably cold target dose to achieve an excessively low rectum dose---without having to specify constraints manually. In other words, the Pareto optimal solutions to the MCO problem, and thus also the range covered by the tilted predictions, should span as exactly as possible the plans which are achievable as well as clinically acceptable. 

Noting that the predictive distributions of the dose statistics have infinite support, we want the dose statistic tilting to direct attention toward the distribution tails. For this, we will exploit the mixture-Gaussian form of the distributions and use exponential tilting, inspired by importance sampling techniques. Given a probability distribution $p(y)$ for $y$, the $\zeta$-exponentially tilted distribution $p^{\zeta}(y)$ is obtained by a change of probability measure according to $p^{\zeta}(y) \propto e^{\zeta^{\operatorname{T}} y} p(y)$. In our case, it can be shown that the tilted predictive distribution is again a Gaussian mixture 
\begin{align*}
&p^{\zeta}(y^* \mid x^*, \{(x^n, y^n)\}_n, \theta) \\
&\quad\quad = \sum_c \frac{\pi_c e^{\zeta^{\operatorname{T}} P \mu_c + \zeta^{\operatorname{T}} P \Sigma_c P^{\operatorname{T}} \zeta / 2}}{\sum_{c'} \pi_{c'} e^{\zeta^{\operatorname{T}} P \mu_{c'} + \zeta^{\operatorname{T}} P \Sigma_{c'} P^{\operatorname{T}} \zeta / 2}} \operatorname{N}(y^* \mid \overline{y} + P \mu_c + P \Sigma_c P^{\operatorname{T}} \zeta, P \Sigma_c P^{\operatorname{T}}).
\end{align*}
An interpretation of this is that the moment-generating function $\operatorname{\mathbb{E}}^{\operatorname{N}(y^* \; \mid \; \overline{y} + P\mu_c, P \Sigma_c P^{\operatorname{T}})} e^{\zeta^{\operatorname{T}} y^*}$ of $y^*$ under $y^* \sim \operatorname{N}(\overline{y} + P\mu_c, P \Sigma_c P^{\operatorname{T}})$ is multiplied to each original mixture weight $\pi_c$ when tilting---that is, more mass in the tilted distribution is assigned to those $y^*$ which are more parallel to $\zeta$. Moreover, the mean shift for each expert class $c$ will be $P \Sigma_c P^{\operatorname{T}}$. This suggests that $\zeta$ should be parallel to the ideal direction of $y^*$ and of appropriate magnitude. It turns out that the main principal component $p_1$ is a good choice for this, with eventual sign modifications for targets in order for distributions of dose statistics to be maximized to be tilted upward, and vice versa. Using the sign convention that $p_1$ should be in the mostly positive direction and accounting for the fact that $\zeta$ should be in units inverse to $y^*$, we may write $\zeta$ as
\begin{equation}
\label{tiltformula}
\zeta = - \frac{\iota}{\sqrt{\sum_c \pi_c (\Sigma_{c, 11} + \mu_{c, 1}^2) - \left( \sum_c \pi_c \mu_{c, 1} \right)^2}} p_1,
\end{equation}
where $\iota \geq 0$ is a constant and the denominator is the predictive standard deviation of $y^*$ along the direction $p_1$. Using this, we may achieve good choices of $\zeta$ only by tuning a single parameter $\iota$. Note that $\zeta = 0$ corresponds to the neutral model. Figure \ref{dvhstilted} shows examples of tilted distributions in comparison to their neutral counterparts.

Finally, to also obtain spatial dose predictions more compatible with the tilted dose statistic predictions, we need to modify the neutral dose prediction model. This is done by retraining the neutral model using the same spatial loss function but with a different DVH loss. Denoting by $\mu^{\zeta}(y)$ the $\zeta$-exponentially tilted predictive mean of $y$, for each tilting $\zeta$, we may obtain a retrained dose prediction model $f^{\zeta}$ by minimizing the same total loss as described in Section \ref{doseprediction} but with $L$ replaced by
\begin{equation}
\label{modifieddvhloss}
L(d, \hat{d}) = L_{\operatorname{spat}}(d, \hat{d}) + L_{\operatorname{DVH}}\!\left( (\psi_j(d))_j, \mu^{\zeta}(\hat{y}) \right)
\end{equation}
---that is, the reference levels for the dose statistics are replaced by the mean of the tilted predictive distribution according to the mixture-of-experts model. 

\subsection{MCO}

Equipped with a method of tilting predictions of both DVHs and spatial dose distributions, we are now set to construct the MCO problem. Unlike in conventional MCO, where each tradeoff should represent a naive ideal---e.g., max-dose functions at zero dose for OARs---we may utilize the prior information contained in the predictive models to design tradeoff objectives in such a way that penalties become relatively small beyond the range of realistic values. To convert the predictions to objectives, we will use the binary cross-entropy objectives presented in \citet{nilsson} and \citet{zhang_dvhpred}. Recall that the marginal predictive distributions for the dose in each voxel $i$ and the value of each dose statistic $j$ are Gaussian and mixture-Gaussian, respectively, following the assumptions on the regression error $\epsilon$ and properties of the mixture-of-experts model, and let the associated univariate cumulative distribution functions be denoted by $F_i^{\zeta}$ and $F_j^{\zeta}$ under a tilting $\zeta$. For each voxel $i$ achieving dose $d_i$ during optimization, the penalty contribution will be based on the binary cross-entropy 
\[
-t_i \log F_i^{\zeta}(d_i) - (1 - t_i) \log(1 - F_i^{\zeta}(d_i))
,\]
where $t_i$ is a binary label set to $1$ if $d_i$ is to be maximized and $0$ otherwise---the contribution from achieving the value $\psi_j(d)$ in dose statistic $j$ is calculated analogously, $t_j$ being the corresponding label. Such an expression has the property that the higher the certainty that the voxel dose or dose statistic value is around some range of values, the more will the objective penalize deviations from those values. At the same time, since the $F_i$ and $F_j$ are continuous and strictly increasing, the optimizer will always have incentive to improve even when beyond the range of typical values. For more details on such objectives---in particular, including an illustration of their behavior for different probability distributions---see \citet{zhang_dvhpred}.

Hence, for each tradeoff, all of the voxels and dose statistics will be included, but with their respective distributions tilted differently. If $\zeta$ is a particular tilting, the associated tradeoff objective $\psi^{\zeta}$ is written as
\begin{equation*}
\label{tradeofffunction}
\begin{split}
\psi^{\zeta}(d) &= w_{\operatorname{spat}} \sum_i r_i \left( -t_i \log F_i^{\zeta}(d_i) - (1 - t_i) \log(1 - F_i^{\zeta}(d_i)) \right) \\
&\quad\quad + \sum_{j \in S_{\operatorname{obj}}} w_j \left( -t_j \log F_j^{\zeta}(\psi_j(d)) -(1 - t_j) \log(1 - F_j^{\zeta}(\psi_j(d))) \right),
\end{split}
\end{equation*}
where the $r_i$ are the relative volumes of the voxels with respect to the outline ROI, summing to unity, $w_{\operatorname{spat}}$ and the $w_j$ are weights and $S_{\operatorname{obj}}$ is a subset of all dose statistics considered. With a set $Z \ni 0$ of appropriately constructed tiltings, where the zero tilt $\zeta = 0$ represents the neutral model, the resulting MCO problem is written as
\begin{equation}
\label{mcoformulation}
\underrel{\text{minimize}}{\eta \in \mathcal{E}} \quad \left\{\psi^{\zeta}(d(\eta))\right\}_{\zeta \in Z},
\end{equation}
where $\eta$ denotes the optimization parameters, $\mathcal{E}$ its feasible set and the dose deposition is assumed to be defined through some mapping $d = d(\eta)$, the details of which depend on the type of optimization used. One example is in fluence map optimization, where $\eta \geq 0$ may represent discretized nonnegative fluences and where $d$ is linear in $\eta$ \citep{ehrgott}; other examples include direct machine parameter optimization for techniques such as dynamic multileaf collimator delivery, sliding-window volumetric modulated arc therapy (VMAT), intensity-modulated proton therapy, tomotherapy and ordinary VMAT \citep{craft, unkelbach}, all but the last-mentioned also with the property of $d$ being linear (or approximately linear) in the optimization variables $\eta$. Standard algorithms exist for numerically solving general MCO problems---for details, see \citet{breedveld} and \citet{rasmus}. For example, one may optimize on the weighted-sum total objective $\sum_{\zeta \in Z} w^{\zeta} \psi^{\zeta}$ using different weight patterns $(w^{\zeta})_{\zeta \in Z}$---specifically, such an algorithm produces $|Z|$ plans using one-hot vectors as weight pattern, called anchor plans, and one using the all-one vector, called the balance plan \citep{craft}. In the end, regardless of the Pareto plan generation algorithm chosen, we obtain a set $\{d_{\operatorname{Pareto}}^k\}_k$ of Pareto optimal plans with corresponding optimization parameters $\{\eta_{\operatorname{Pareto}}^k\}_k$.

To decide upon a final plan, the user now has the opportunity to intervene by navigating on the surface of Pareto plans, which numerically amounts to choosing an interpolation $\sum_k \lambda_k d_{\operatorname{Pareto}}^k$ with nonnegative coefficients $\lambda = (\lambda_k)_k$ of unit sum. In a treatment planning system such as RayStation (RaySearch Laboratories, Stockholm, Sweden), for instance, one can use sliders to balance the tradeoffs while inspecting in real time the spatial dose, DVHs and clinical goals of the navigated plan. For delivery techniques equipped with the abovementioned dose--variable linearity property, the analogous interpolation $\sum_k \lambda_k \eta_{\operatorname{Pareto}}^k$ constitutes a feasible plan (if the feasible set $\mathcal{E}$ is convex) that recreates the navigated dose distribution, meaning that any navigated plan is directly realizable---otherwise, an additional optimization is performed to mimic the navigated plan using feasible optimization variables \citep{bokrantz_first, craft}. To the extent that the navigated dose may guaranteed to be physically realistic, this motivates the generation of Pareto plans and associated optimization variables rather than directly performing navigation on the predicted doses. Alternatively to manual navigation, in RayStation, it is also possible to automatically navigate by specifying some objectives and constraints summarizing one's preferences, e.g. based on clinical goals. In particular, dividing a set of clinical goals $\{\pm \psi_j(d) \leq \pm \hat{\psi}_j\}_{j \in S_{\operatorname{CG}}}$, signs depending on $t_j$, into objectives and constraints $S_{\operatorname{CG}} = S_{\operatorname{CG, \, obj}} \cup S_{\operatorname{CG, \, constr}}$, we may write the autonavigation optimization problem as
\begin{equation}
\label{autonavigation}
\begin{aligned}
& \!\!\!\!\! \underset{\lambda \; : \; \lambda \geq 0, \; 1^{\operatorname{T}}\lambda = 1}{\text{minimize}}
&& \sum_{j \in S_{\operatorname{CG,\, obj}}} (\psi_j(d) - \hat{\psi}_j)_{\pm}^2 \\
& \text{subject to} 
&& \quad \,\, d = \sum_k \lambda_k d_{\operatorname{Pareto}}^k, \\
&&& \quad \pm \psi_j(d) \leq \pm \, \hat{\psi}_j \quad \text{for all } j \in S_{\operatorname{CG, \, constr}},
\end{aligned}
\end{equation}
$(x)_{\pm}$ denoting the positive or negative part of $x$. While autonavigation has the potential of making the proposed semiautomatic workflow fully automatic, due to the fact that the construction of (\ref{autonavigation}) needs to be based on domain knowledge, it may also be viewed an enhanced manual step.

\subsection{Computational study}

To demonstrate the proposed semiautomatic pipeline, we performed numerical experiments using a dataset originating from Iridium Cancer Network (Antwerp, Belgium), which comprised $91$ retrospective treatment plans of prostate cancer patients having undergone a prostatectomy prior to radiation therapy. The patients were treated with dual $360$-degree VMAT arcs and prescriptions of $7000 \; \mathrm{cGy}$ in the prostate bed and $5600 \; \mathrm{cGy}$ in the seminal vesicles and pelvic nodes, with final doses calculated by the collapsed cone algorithm in RayStation. The dataset was split into $84$ training or validation patients and $7$ test patients---the models trained using the former set were used to make predictions and create Pareto plans for each patient in the latter. For all parts of the numerical experiments, the set $\mathcal{R}$ of ROIs considered were the prostate PTV, seminal vesicles PTV, rectum and bladder regions, of which the former two constituted $\mathcal{R}_{\operatorname{target}}$.

\begin{figure}[h]
\centering
\includegraphics[width=\textwidth]{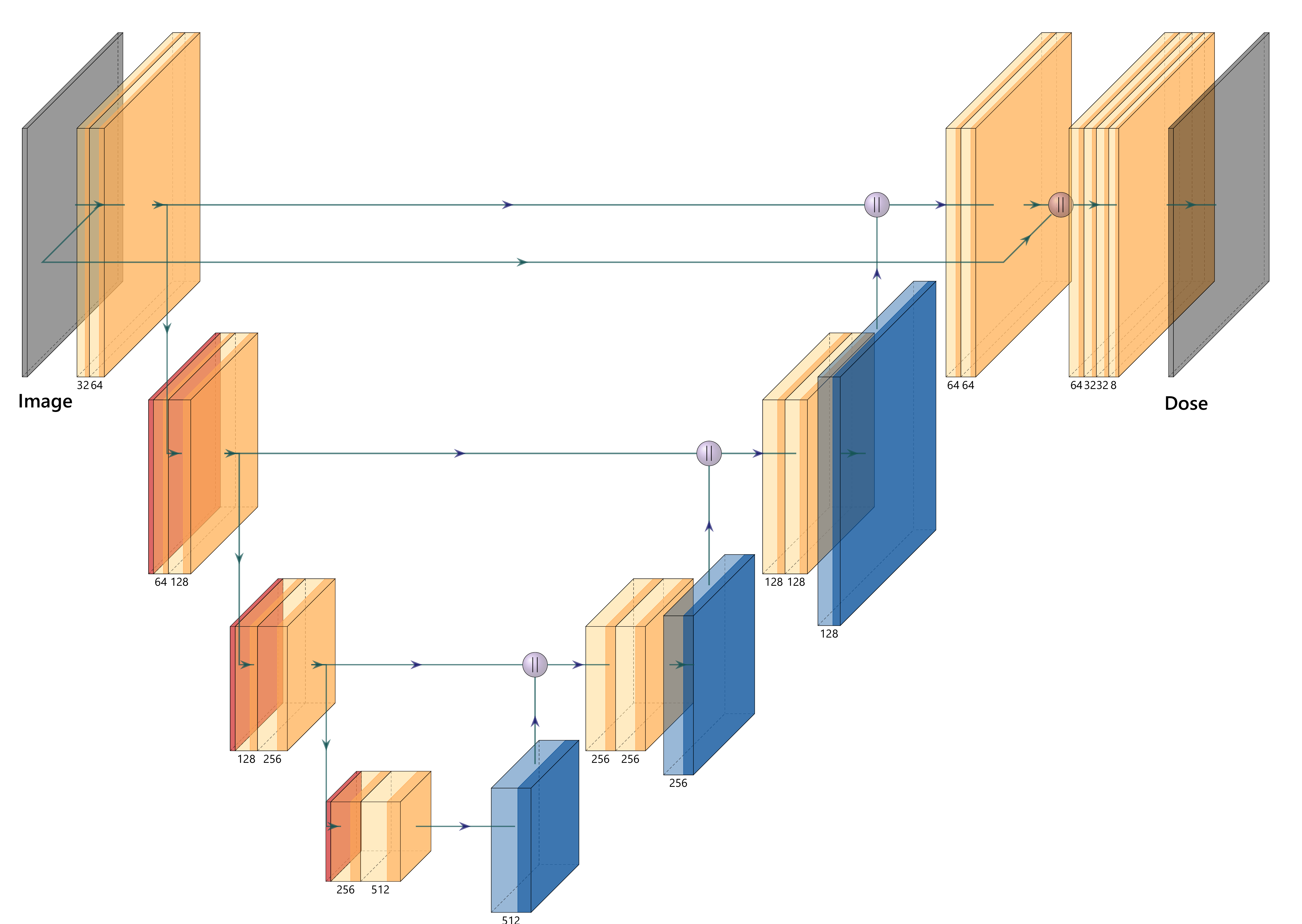}
\caption{Illustration of the architecture of the U-net used for neutral and tilted spatial dose predictions. The encoder part consists of convolutional (yellow) and max-pooling (red) layers, and the decoder upsamples using transpose-convolutional (blue) layers. After each layer, an activation is applied if indicated with a darker color, which is followed by layer normalization.}
\label{architecture}
\end{figure}

The neural network used for the neutral and tilted spatial dose prediction models was implemented in TensorFlow 2.4, using the three-dimensional U-net architecture depicted in Figure \ref{architecture}. Convolutional and max-pooling layers were used for the encoder part and transpose-convolutional layers for the decoder part, along with a sequence of convolutional layers before the final output is produced, and layer normalization was applied after each rectified linear unit (ReLU) activation. The input images were preprocessed to arrays of size $64 \times 48 \times 83 \times 6$ with a $5 \; \mathrm{mm}$ voxel resolution, where, as expanded in (\ref{voxelinfo}), the $6$-dimensional information $x_i$ for each voxel $i$ is the binary ROI encoding concatenated with the target distance transforms. Comprising around $2 \cdot 10^7$ weights, the models were fitted using a standard Adam optimizer and a training--validation split of $75$ and $9$ patients, respectively. The gradient term weight $\alpha$ in $L_{\operatorname{spat}}$ was set to $1$, and the dose statistics used in the DVH-based loss $L_{\operatorname{DVH}}$ were dose-at-volume functions $\{\operatorname{D}_{i / 100}\}_{i=1}^{99}$ at each integer-percentage volume level from $1 \; \%$ to $99 \; \%$, with individual dose statistic weights $\sigma_j^{-2}$ manually tuned such that targets had total weight $10$ times that of OARs. For fitting the neutral model, the training lasted around $100$ epochs, after which an approximate minimum in validation loss was reached. The tilted models were obtained by retraining the neutral network for another $10$ epochs using the modified DVH loss (\ref{modifieddvhloss}), the length of the training set so as not to deviate too far from the neutral prediction while still achieving the desired tilted effect.

The set $\{\psi_j\}_j$ of all dose statistics to be considered was dose-at-volume functions $\{\operatorname{D}_{i / 100}\}_{i=1}^{99}$ for all four ROIs, along with additional mean-tail-dose functions $\{\operatorname{MTD}_{i / 100}^{\pm}\}_{i=1}^{99}$ \citep{romeijn} for targets (using lower mean-tail-dose $\operatorname{MTD}^-_v$ for $v \geq 0.5$ and upper mean-tail-dose $\operatorname{MTD}^+_v$ for $v < 0.5$) to increase control of distribution tails. In order to achieve better predictive accuracy given the small dataset, we used separate models for each ROI at the cost of sacrificing inter-ROI dependencies. The raw inputs $\phi_{\operatorname{tot}}(x)$ were constructed according to (\ref{dspredinput}), where the geometric inputs $\phi_{\operatorname{geom}}(x)$ were distance transforms from each ROI in $\mathcal{R}_{\operatorname{OAR}}$ to each ROI in $\mathcal{R}$ represented by histograms in a total of $88$ dimensions. Concatenating with the dose statistic values predicted by the spatial model and using the isomap algorithm, embeddings $\phi_{\operatorname{iso}}(x)$ were reduced to $10$ dimensions. Meanwhile, the $4$ largest principal components were used in $P$, leading to inputs $\phi_{\operatorname{iso}}(x)$ and outputs $y_{\operatorname{pc}}$ of the mixture-of-experts models being of dimension $10$ and $4$, respectively. Each mixture-of-experts model used $C = 16$ classes with a posterior sample size of $10$ for each mean--covariance pair $(\mu_c, \Sigma_c)$. The set $Z$ of tilts was constructed by using (\ref{tiltformula}) on each ROI in $\mathcal{R}$ at a time while letting predictions for other ROIs remain neutral, with the constant $\iota$ tuned manually, resulting in $|Z| = |\mathcal{R}| + 1 = 5$ tilts in total including the original all-neutral model.

Constructing tradeoff objectives from the neutral and tilted predictions according to (\ref{tradeofffunction}), where the subset $S_{\operatorname{obj}}$ was chosen to be dose-at-volume statistics at volume levels $10 \; \%, 20 \; \%, \dots, 99 \; \%$ for OARs and both dose-at-volume and mean-tail-dose statistics at $1 \; \%, 2\; \%, 5\; \%, 10 \; \%, 20 \; \%, \dots, 90 \; \%, 95 \; \%, 98 \; \%, 99 \; \%$, the MCO formulation (\ref{mcoformulation}) was implemented and solved numerically in a research version of RayStation 11A. A standard algorithm \citep{rasmus} was used to generate $|Z| + 1 = 6$ fluence-based Pareto plans---five anchor plans and one balance plan---each using $40$ iterations of the in-house sequential quadratic programming solver. To evaluate the proposed methodology against a single-plan-output automated planning algorithm, we compared the neutral anchor plan to the best possible plan in the navigation space, here represented by the autonavigated plan using the optimization formulation (\ref{autonavigation}) and the clinical goals shown in Table \ref{tab:autonavigationgoals}. Note that the use of autonavigation is mostly intended to serve as a means of evaluate the quality of the produced Pareto frontier approximation, replacing a human user for the purposes of this study. We also compared OAR sparing in the clinical plan to the deliverable VMAT plan constructed from the best navigated plan using the in-built conversion process in RayStation. 

\begin{table}[h]
\caption{\label{tab:autonavigationgoals} Clinical goals used in the autonavigation optimization problem (\ref{autonavigation}), either as constraints or objectives.}
\centering
\begin{tabular}{lll}
\toprule
ROI & Goal & Group \\
\midrule
PTV, prostate & $\operatorname{D}_{98 \, \%} \geq 6650 \; \mathrm{cGy}$ & Constraint \\
PTV, prostate & $\operatorname{D}_{2 \, \%} \leq 7300 \; \mathrm{cGy}$ & Constraint \\
PTV, seminal vesicles & $\operatorname{D}_{98 \, \%} \geq 5250 \; \mathrm{cGy}$ & Constraint \\
PTV, seminal vesicles & $\operatorname{D}_{5 \, \%} \leq 5850 \; \mathrm{cGy}$ & Constraint \\
Rectum & Minimize mean dose as much as possible & Objective \\
Bladder & Minimize mean dose as much as possible & Objective \\
\bottomrule
\end{tabular}
\label{autonavigationgoals}
\end{table}

\section{Results}

\begin{figure}[h]
\centering
\begin{subfigure}[t]{\textwidth}
\centering
\includegraphics[width=\textwidth]{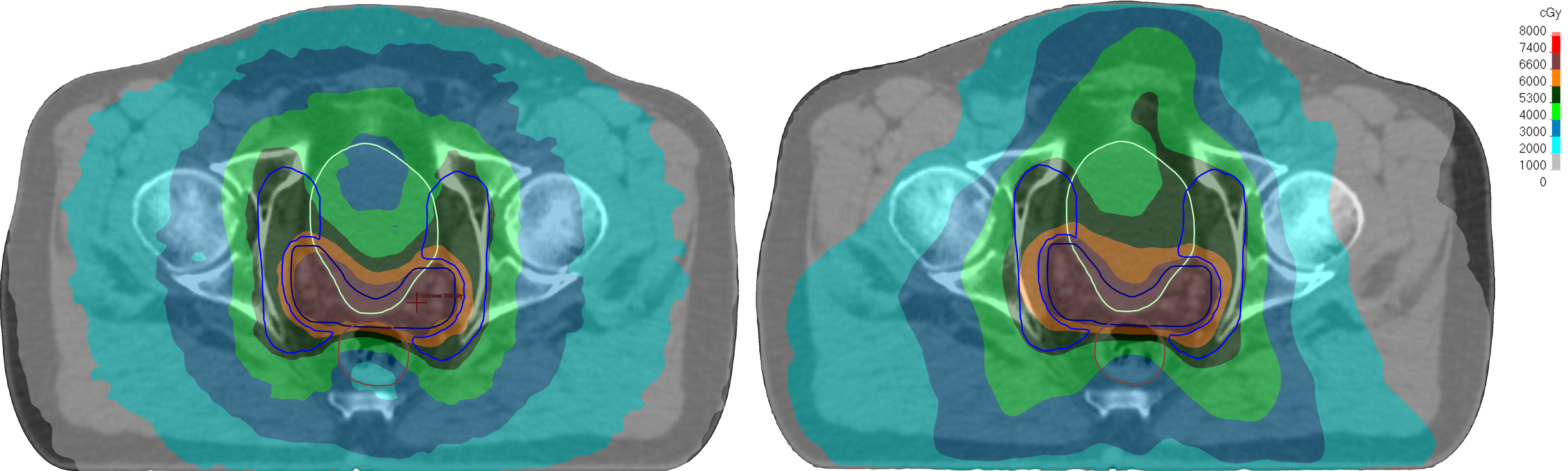}
\caption{}
\end{subfigure}%
\vspace{0.5cm}
\begin{subfigure}[b]{0.7\textwidth}
\centering
\includegraphics[width=\textwidth]{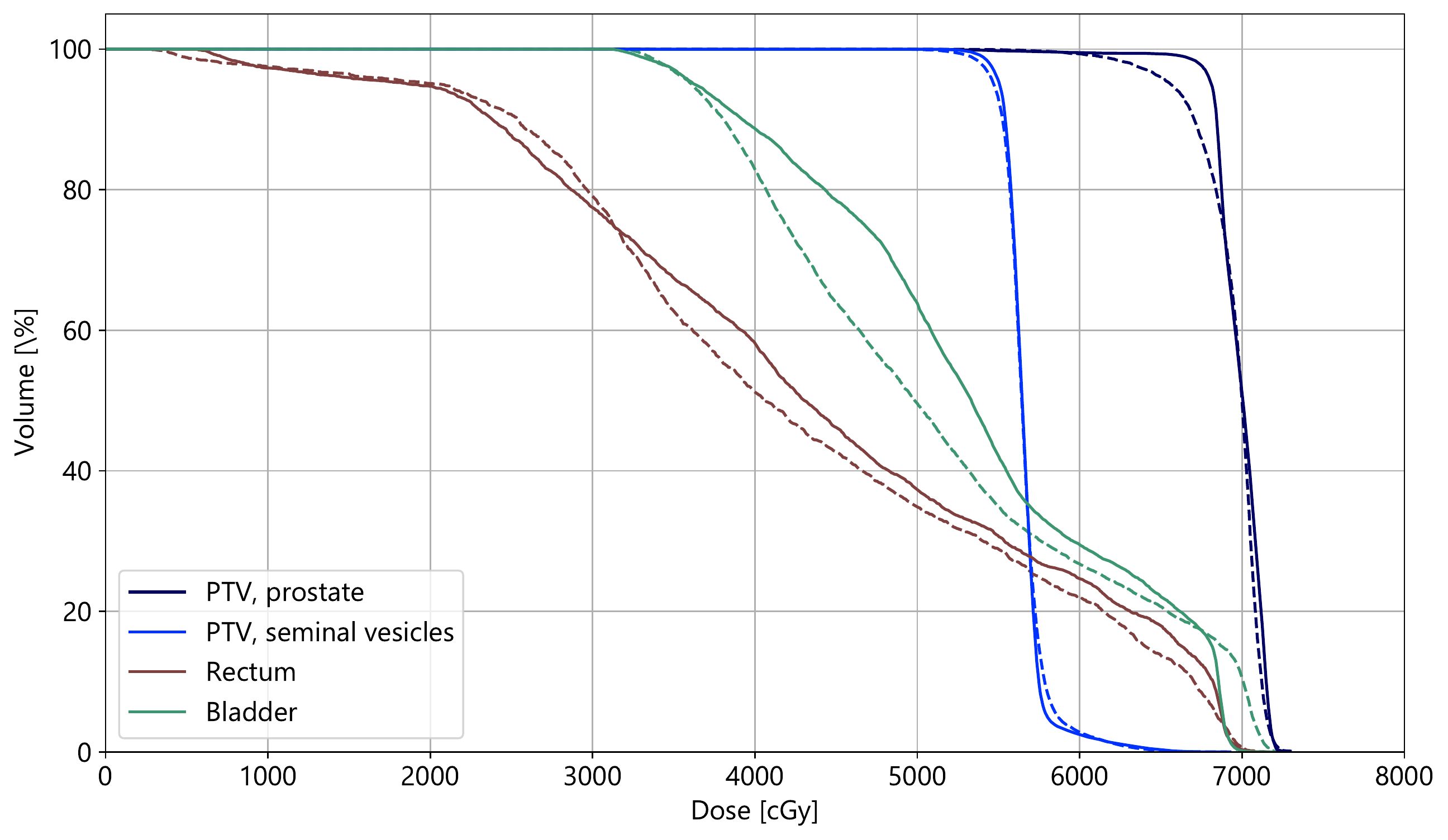}
\caption{}
\end{subfigure}
\caption{(a) Transversal cuts of the neutrally predicted spatial dose (left) and the clinical ground truth counterpart (right). (b) Comparison in DVH between those of the neutrally predicted spatial dose (dashed) and the clinical ground truth (solid).}
\label{neutralvsground}
\end{figure}

\begin{figure}[h]
\centering
\includegraphics[width=\textwidth]{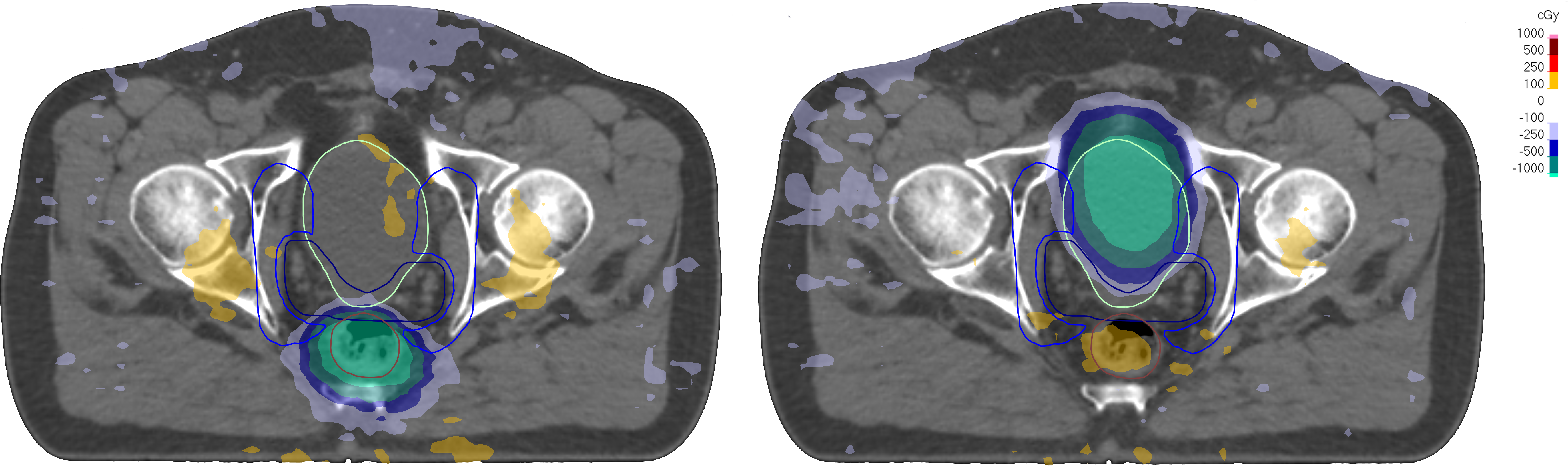}
\caption{Difference maps for the rectum- and bladder-tilted spatial dose predictions (left and right, respectively) versus the neutral prediction.}
\label{diffstilted}
\end{figure}

\begin{figure}[h]
\centering
\includegraphics[width=\textwidth]{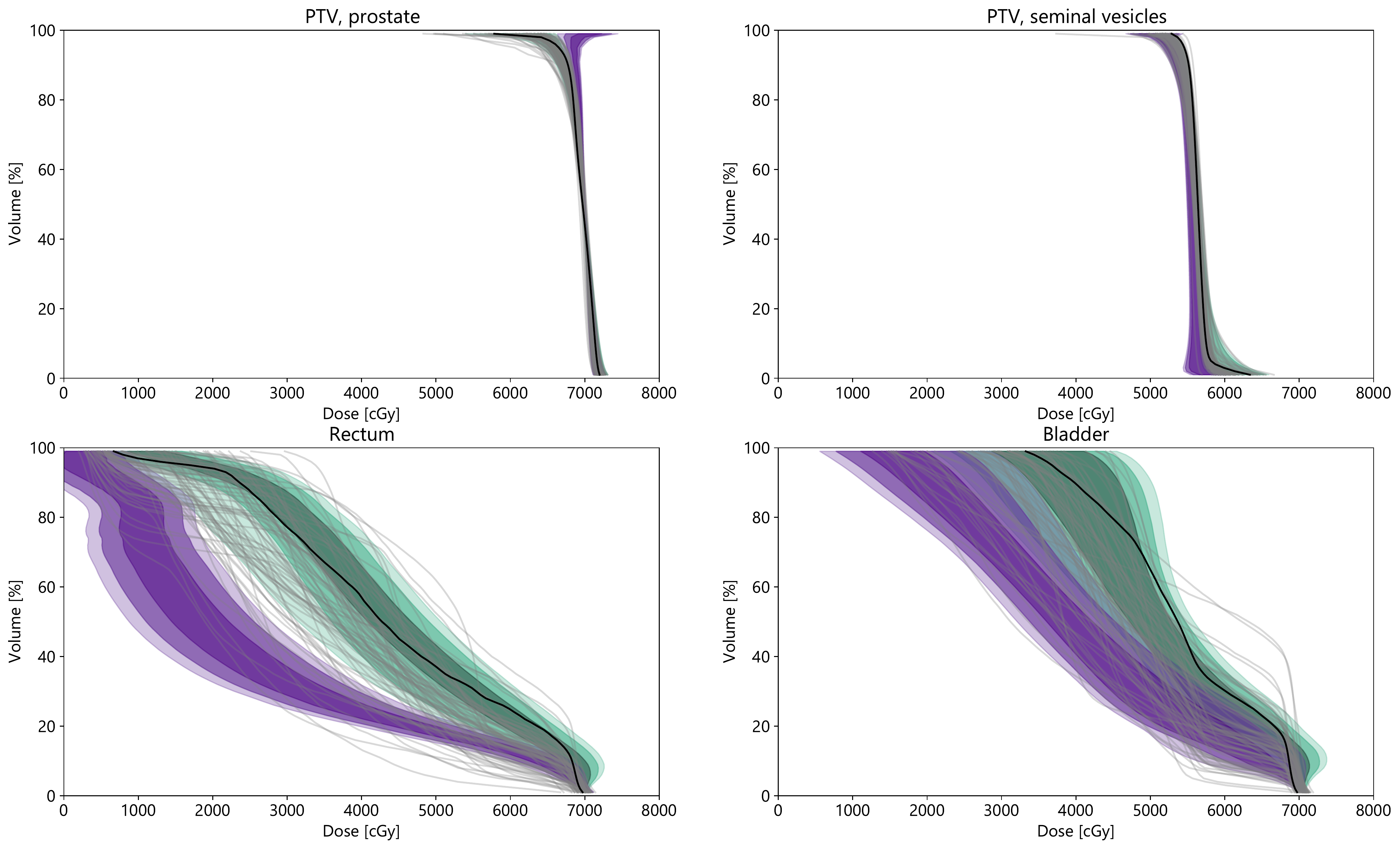}
\caption{Illustration of neutral (green) versus tilted (purple) dose statistic predictions, where the shaded bands correspond to $99 \; \%$, $95 \; \%$ and $70 \; \%$ prediction intervals for each dose-at-volume statistic $\operatorname{D}_v$. The training DVHs are shown in gray and the ground truth in black. Note that since the prediction intervals are displayed pointwise for each $v$, there is no guarantee that the interval limits are monotonous in $v$.}
\label{dvhstilted}
\end{figure}

Starting with the prediction models, although they are not the main focus of our numerical experiments, we may still validate qualitatively that their purposes are fulfilled. Figure \ref{neutralvsground} shows a comparison between the neutrally predicted spatial dose and the corresponding ground truth for a patient in the test dataset. While some noticeable differences in both spatial dose and DVH are present---for example, the predicted dose decays faster beyond the targets and has inferior target coverage in the prostate PTV---the overall quality of the dose prediction is quite sufficient for being a starting point in our pipeline. Furthermore, Figure \ref{diffstilted} shows difference maps between the neutral and the rectum- and bladder-tilted models. One can see that while the main differences are located in the respective ROIs, the dose shifts are relatively smooth, extending beyond the ROIs subject to tilting, and also smaller in or near overlap regions with the targets, as would be expected of a physically realistic dose. These properties are indicative of the Sobolev and DVH loss functions achieving the desired effects of penalizing spatial unevenness and excessive DVH deviation, respectively, in combination with the natural smoothness implied by the convolutional neural network architecture. As for the dose statistic prediction, the dimensionality-normalized root mean squared error 
\[
\sqrt{\frac{1}{\operatorname{dim} \mathcal{Y}}\operatorname{\mathbb{E}}^{p(y \; \mid \; x^*, \{(x^n, y^n)\}_n)} |y - y^*|^2}
\]
had means $137 \; \operatorname{cGy}$, $109 \; \operatorname{cGy}$, $722 \; \operatorname{cGy}$, $412 \; \operatorname{cGy}$ and standard deviations $21 \; \operatorname{cGy}$, $13 \; \operatorname{cGy}$, $297 \; \operatorname{cGy}$, $111 \; \operatorname{cGy}$ for the prostate PTV, seminal vesicles PTV, rectum and bladder ROIs, respectively, over the seven test patients. Moreover, Figure \ref{dvhstilted} shows pointwise DVH confidence bands for both the neutral and the tilted model over the four ROIs considered for the same test patient. One can see that the ground truth DVHs are well within the neutral prediction bands and that the tilted counterparts have predicted significantly more uniform doses to the targets and lower doses to the OARs. Hence, this assures us that the dose statistic prediction is reasonable and that the tilted models indeed correspond to more optimistic predictions.

\begin{figure}[h]
\centering
\includegraphics[width=\textwidth]{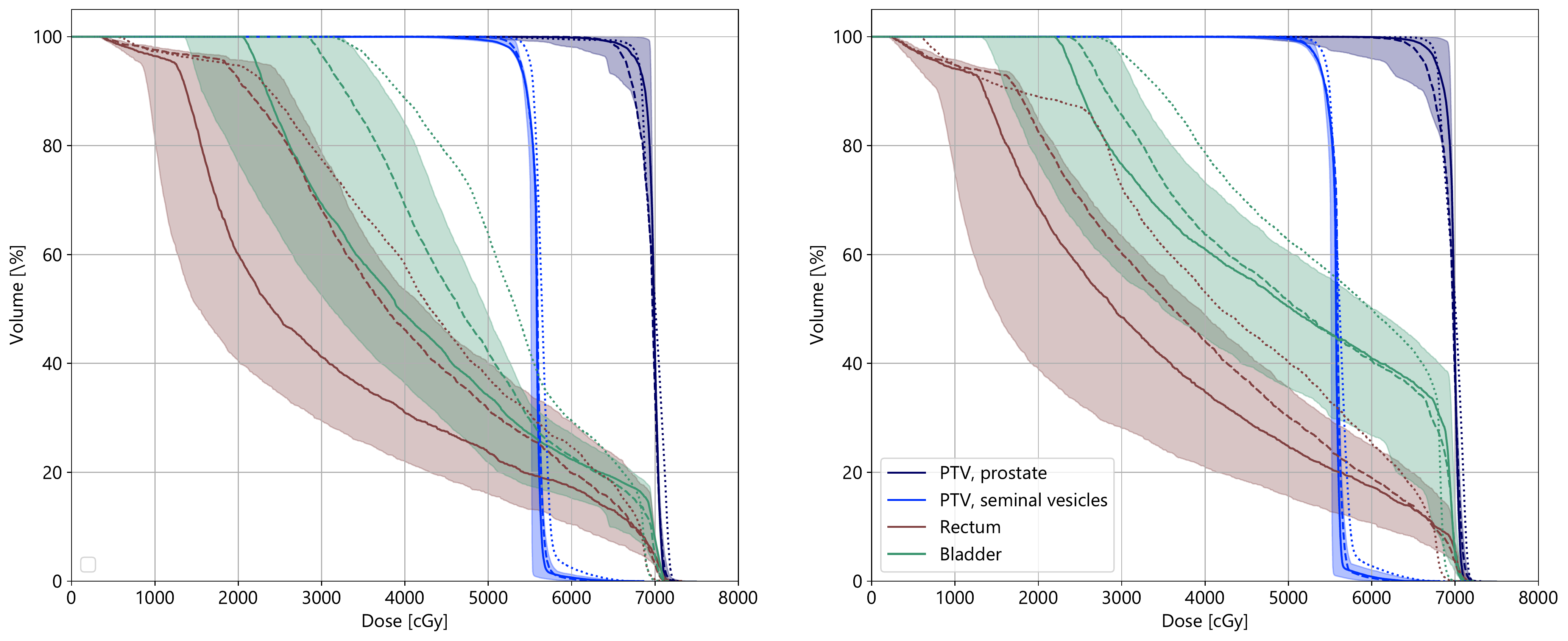}
\caption{Visualization of the DVH spans of the produced Pareto plans for two test patients, where the shaded regions lie between minima and maxima for each dose-at-volume statistic $\operatorname{D}_v$, pointwise for each $v$. The lines show the DVHs for the neutral anchor (dashed), the autonavigated (solid) and the clinical (dotted) plan.}
\label{dvhspan}
\end{figure}

\begin{figure}[h]
\centering
\begin{subfigure}[c]{0.5\textwidth}
\centering
\includegraphics[width=\textwidth]{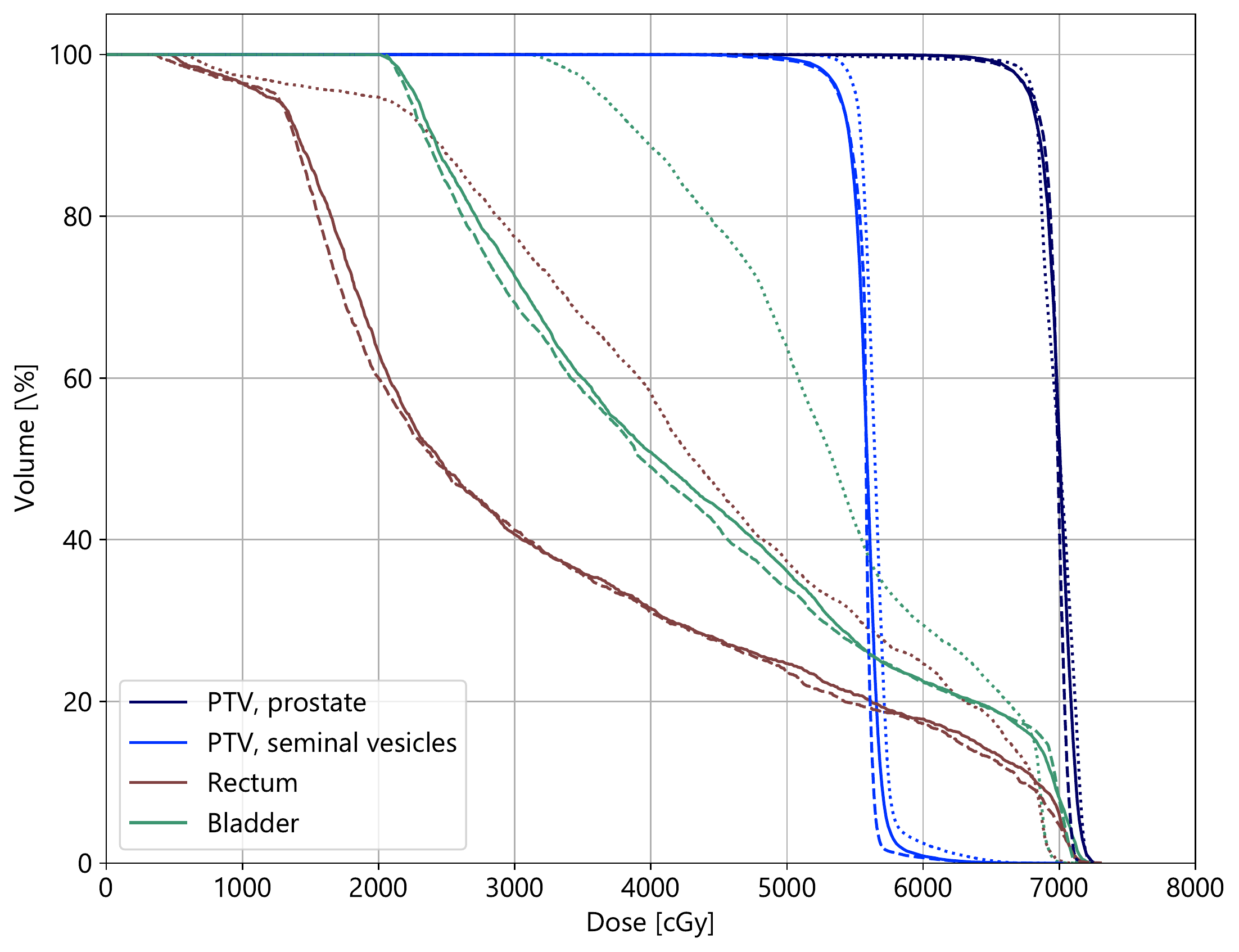}
\caption{}
\end{subfigure}%
\begin{subfigure}[c]{0.5\textwidth}
\centering
\includegraphics[width=\textwidth]{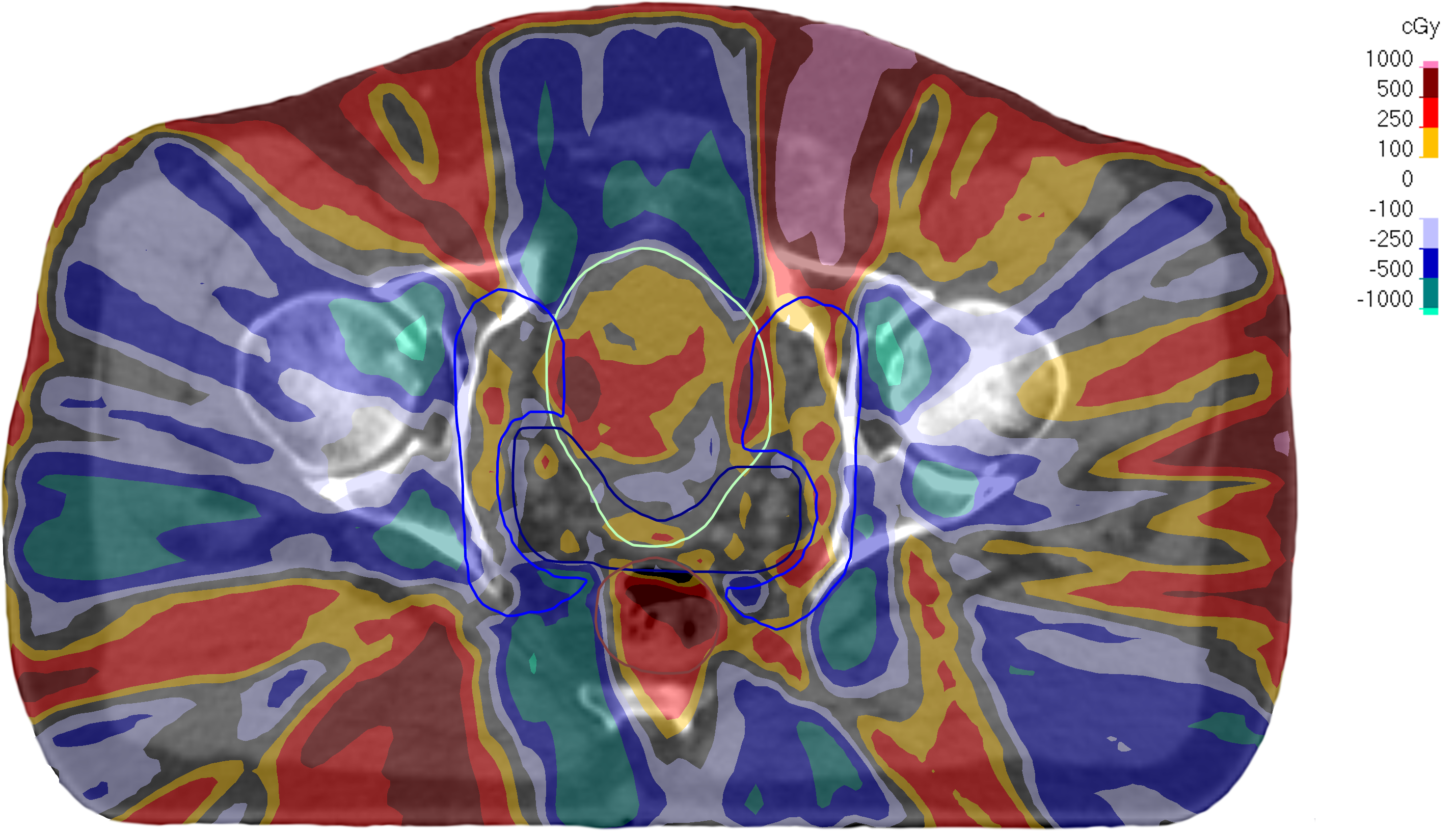}
\caption{}
\end{subfigure}
\caption{(a) Comparison in DVH between those of the fluence-based autonavigated plan (dashed), the converted machine parameter--based deliverable plan (solid) and the clinical plan (dotted). (b) Difference map between the deliverable and the autonavigated plans.}
\label{deliverablevsground}
\end{figure}

For each patient, obtaining the six Pareto plans took around five minutes. To visualize the range of plans achievable with the produced approximation of the Pareto set, Figure \ref{dvhspan} shows the span of DVHs achievable for each ROI on the test patient alongside the fluence-based neutral and navigated plans, where the navigation has been performed by the autonavigation algorithm. One can see that the fluence-based navigated plan, representing for our purposes the best plan in the Pareto set, has better sparing of the rectum and bladder as well as better target coverage compared to the neutral anchor plan, which represents what would have been the output had we pursued a single-plan approach. As may be seen in Figures \ref{neutralvsground} and \ref{dvhstilted}, the predictions for the prostate PTV are slightly too cold, which has indeed translated to a similar effect in the neutral anchor plan. On the other hand, by navigating closer to the prostate PTV anchor plan, we are able to improve the target coverage---analogously, we are able to simultaneously achieve better OAR sparing. This illustrates, in particular, the necessity for a means of easily adjusting the preliminary output of an automated treatment planning algorithm, for which the MCO framework is well-adapted. Furthermore, Table \ref{tab:autonavigationresults} shows the average differences in dose statistic values between the fluence-based navigated and neutral plans and between the machine parameter--based navigated and clinical plans. Here, it is evident that the fluence-based and machine parameter--based navigated plans present improvements over the neutral and clinical plans, respectively. One can also see in Figure \ref{deliverablevsground} that the machine parameter conversion is able to keep differences between the machine parameter--based and fluence-based plans small, which reassures us that the fluence-based MCO, despite being an idealization, is physically realistic to a certain extent. Although similar experiments need to be performed on larger datasets and other treatment sites in order to be able to draw definitive conclusions, all in all, the results show that our proposed semiautomatic methodology has the potential of improving the quality of plans produced by fully automated pipelines while offering the additional flexibility of real-time MCO navigation.

\begin{table}[h]
\caption{\label{tab:autonavigationresults}Differences in dose statistic values between the autonavigated (Nav) and the neutral anchor (Neu) plan, and between the deliverable (Del) and the clinical (Clin) plan. The displayed values are sample means and standard deviations of the pairwise differences over the test dataset of seven patients.}
\centering
\begin{tabular}{llrr}
\toprule
ROI & Dose statistic & $\text{Nav} - \text{Neu}$ ($\mathrm{cGy}$) & $\text{Del} - \text{Clin}$ ($\mathrm{cGy}$) \\
\midrule
PTV, prostate & $\operatorname{D}_{98 \, \%}$ & $88 \pm 46$ & $57 \pm 118$ \\
PTV, prostate & $\operatorname{D}_{2 \, \%}$ & $-9 \pm 19$ & $-35 \pm 52$ \\
PTV, seminal vesicles & $\operatorname{D}_{98 \, \%}$ & $-71 \pm 58$ & $-6 \pm 110$  \\
PTV, seminal vesicles & $\operatorname{D}_{5 \, \%}$ & $-35 \pm 22$ & $-70 \pm 42$ \\
Rectum & Mean dose & $-509 \pm 357$ & $-786 \pm 390$ \\
Bladder & Mean dose & $-617 \pm 365$ & $-1161 \pm 589$ \\
\bottomrule
\end{tabular}
\label{autonavigationresults}
\end{table}

\begin{figure}[h]
\centering
\includegraphics[width=\textwidth]{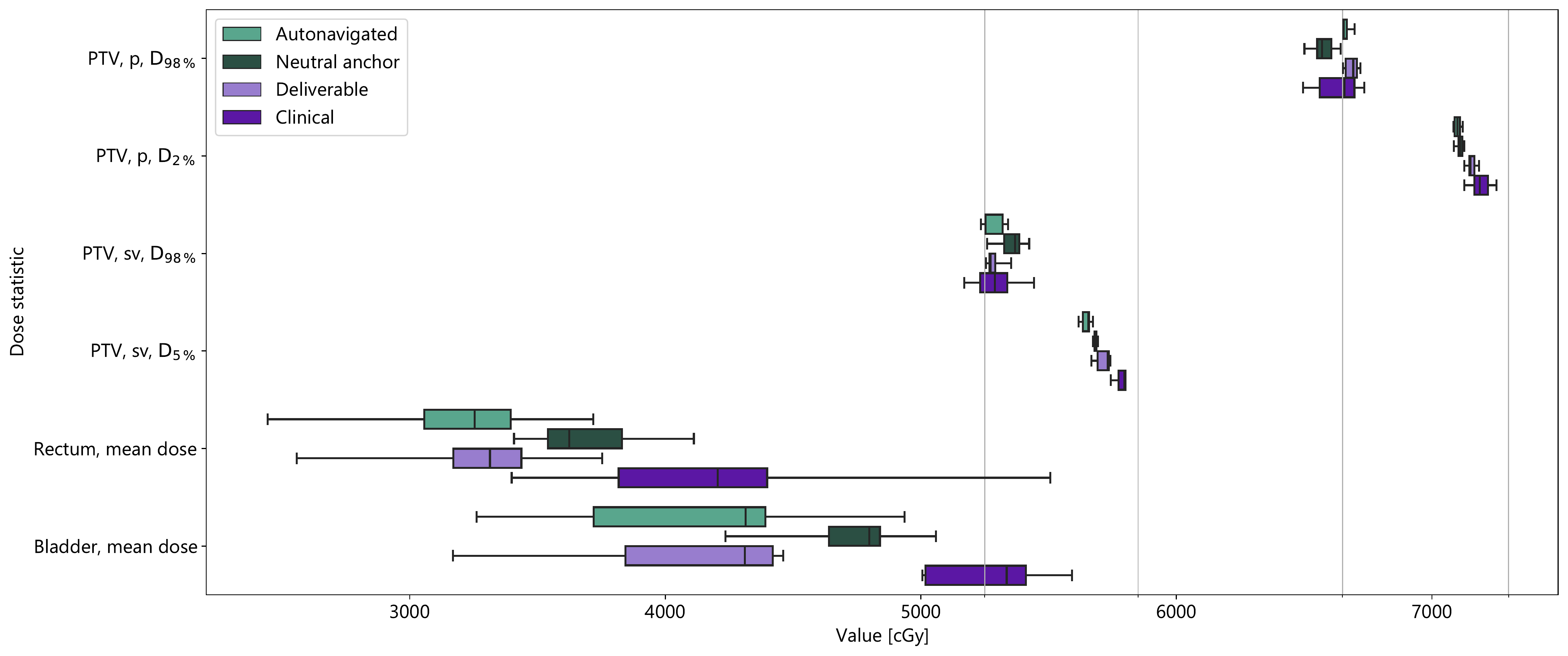}
\caption{Boxplot of the spread of each dose statistic over the seven patients for the autonavigated, neutral anchor, deliverable and clinical plans. The acceptance levels for the clinical goals on the targets are marked with vertical gridlines.}
\label{dosestatisticboxplot}
\end{figure}

\section{Discussion}

Despite easily being able to produce acceptable or near-acceptable plans in most cases, current automated treatment planning algorithms are approaching a plateau in performance as function of training data size and quality---in the last fraction of cases, they struggle with consistently achieving clinical quality without additional post-processing. Instead of trying to make further improvements of such algorithms, we have argued that this is indicative of more fundamental weaknesses of the automated planning paradigm and sought to develop a semiautomatic counterpart, where the output is a range of plans instead of a single plan and the final plan may be decided upon after optional manual adjustments. Suiting our purposes well, the established framework of MCO was combined with machine learning and optimization methods known from classical prediction--mimicking automated planning algorithms. In particular, a three-dimensional convolutional U-net with specially developed loss functions was used to predict spatial dose, and a nonparametric Bayesian regression method was used to estimate multivariate predictive distributions of dose statistics. These neutral predictions were then tilted by a change of probability measure into ones biased toward different groups of planning goals at a time---importantly, this idea allowed us to simulate different goal prioritizations without requiring such plans to be included in the training data. Constructing tradeoffs based on predictive distributions of spatial dose as well as dose statistics, the resulting MCO problem was numerically solved to produce Pareto plans, presenting the user with a navigation-ready setup. Compared to previous literature on the subject, our proposed method differs in that actual fluence-based or machine parameter--based plans are generated, as opposed to only considering predicted doses, and that we have completed the semiautomatic pipeline by performing navigation and converting to a deliverable plan. The results are promising, illustrating the facts that one may often improve the neutral anchor plan by navigating on the Pareto surface and that the corresponding deliverable plan may often be better than the clinical ground truth. 

Among the advantages of our proposed method are the limited demands on the size and quality of the input training data---as the output no longer needs to precisely match the clinical protocol at hand, the data requirements are arguably even less strict than for a conventional automated planning algorithm---the minimal need for tuning and commissioning domain-knowledge objectives in the dose mimicking problem, and the real-time navigation interface facilitating reaching consensus between treatment planners and radiation oncologists. Furthermore, if one prefers a completely automated approach, the manual navigation step may be replaced by an autonavigation algorithm based on clinical goals. While flucence-based VMAT Pareto plans, having the also important advantage of computational speed but the disadvantage of requiring an additional conversion step, were used for the numerical experiments in this paper, there are numerous other common delivery techniques for which the navigated plan is directly or almost directly deliverable, examples including DMLC, sliding-window VMAT, intensity-modulated proton therapy and tomotherapy. On the other hand, our workflow has the additional step of creating tilted versions of the predictions, which is relatively sensitive to the estimated multivariate predictive distribution of the dose statistics and thus includes a certain extent of hand-crafting. 

While the present work primarily focuses on conceptually demonstrating the proposed method, there are many interesting directions for future research. For instance, to further conform to the philosophy of probabilistic machine learning, one may seek to develop a spatial dose prediction method which is able to estimate the multivariate probability distribution over all voxel doses, e.g. using a probabilistic extension of a U-net \citep{kohl}, Gaussian processes \citep{rasmussenwilliams} or Markov random field methods \citep{murphy}. While the Sobolev space--inspired loss function used for spatial dose prediction is shown to work well, one would need further numerical experiments to confirm that it improves the dose prediction quality over ordinary loss functions. Other possibilities include investigating prediction of all dose statistics at once instead of using ROI-specific models and developing a more systematic and customized method of tilting dose statistic predictions, particularly in such a way that one may allocate higher importance to certain dose statistics instead of, as is the case with the current method, treating the whole DVH for each ROI relatively equally. For example, one could investigate ways of constructing more advanced tilting vectors $\zeta$, e.g by analyzing regression residuals such as in \citet{zhangge}. Lastly, another important next step of evaluating the proposed methodology is to include human treatment planners and clinicians to perform the navigation part and evaluate the quality of the end result. Although more evaluations of the method on other datasets and treatment sites remain before definitive conclusions may be drawn regarding how the method compares against existing ones, the semiautomated data-driven workflow presented in this work has shown considerable promise in alleviating the radiation therapy treatment planning process from labor-intensive and monotonous tasks.

\section{Conclusion}

In this work, we have presented a new semiautomatic treatment planning pipeline in which knowledge extracted from historical plans is utilized to generate for each new patient a Pareto set representation, optionally allowing for manual adjustments through navigation before a final plan is settled upon. By unifying ideas from the current automated planning paradigm with MCO---in particular, where neutral spatial dose and dose statistic predictions are augmented with counterparts biased toward different planning goals---we are able to achieve this with minimal quality requirements on input training data. The computational study illustrates a substantial benefit from exploring different tradeoffs compared to a single-plan-output approach and the potential of ultimately obtaining deliverable plans better than the clinical ground truth. For clinics wishing to automate labor-intensive tasks in the treatment planning process while preferring to maintain a certain extent of manual control, the proposed methodology constitutes a viable option to the current fully automated planning paradigm.

\printbibliography

\end{document}